\documentclass[%
 reprint,
 amsmath,amssymb,
 aps,
prl,
superscriptaddress
]{revtex4-2}

\usepackage{amsmath}
\usepackage{graphicx}
\usepackage{dcolumn}
\usepackage{bm}

\usepackage[normalem]{ulem}
\usepackage{color}
\definecolor{myorange}{RGB}{199,146,32}

\begin{document}

\title{Percolating Cosmic String Networks from Kination}

\author{Joseph P. Conlon}
\email{joseph.conlon@physics.ox.ac.uk}
\affiliation{%
 Rudolf Peierls Centre for Theoretical Physics, Beecroft Building, Clarendon Laboratory, Parks Road, Department of Physics, University of Oxford OX1 3PU, UK 
}%
\author{Edmund J. Copeland}%
 \email{ed.copeland@nottingham.ac.uk}
\affiliation{%
 School of Physics and Astronomy,  University of Nottingham, Nottingham, NG7 2RD, UK.
}%
\author{Edward Hardy}
\email{edward.hardy@physics.ox.ac.uk}
\affiliation{%
 Rudolf Peierls Centre for Theoretical Physics, Beecroft Building, Clarendon Laboratory, Parks Road, Department of Physics, University of Oxford OX1 3PU, UK
}%
\author{Noelia Sánchez González}
\email{noelia.sanchezgonzalez@physics.ox.ac.uk}
\affiliation{%
 Rudolf Peierls Centre for Theoretical Physics, Beecroft Building, Clarendon Laboratory, Parks Road, Department of Physics, University of Oxford OX1 3PU, UK
}%
\date{\today}

\begin{abstract}
We describe a new mechanism, whose ingredients are realised in string compactifications, for the formation of cosmic (super)string networks. Oscillating string loops grow when their tension $\mu$ decreases with time. If $2H + \dot{\mu}/\mu < 0$, where $H$ is the Hubble parameter, loops grow faster than the scale factor and an initial population of isolated small loops (for example,  produced by nucleation) can grow, percolate and form a network. This condition is satisfied for fundamental strings in the background of a kinating volume modulus rolling towards the asymptotic large volume region of moduli space. Such long kination epochs are motivated in string cosmology by both the electroweak hierarchy problem and the need to solve the overshoot problem. The tension of such a network today is set by the final vacuum; for phenomenologically appealing Large Volume Scenario (LVS) vacua, this would lead to a fundamental string network with $G \mu \sim 10^{-10}$.

\end{abstract}

\maketitle


\section{\label{sec:level1} Introduction}

Networks of cosmic strings, reviewed in \cite{Vilenkin:2000jqa, Copeland:2009ga}, are a candidate for new physics on cosmological scales. Their signatures include gravitational waves (e.g. as a candidate explanation of the recent NANOGrav results \cite{NANOGrav:2023gor}), CMB anisotropies and direct lensing of distant galaxies. For $G \mu \lesssim 10^{-7}$, where $G$ is Newton's constant and $\mu$ is the string tension, the existence of such networks in the universe is compatible with current observations (the precise limit depends on the type of loop distribution assumed for the string network, e.g. see \cite{LIGOScientific:2021nrg} for details of the current LIGO bounds on Nambu-Goto strings). Strings are topological defects and so are normally assumed to require a symmetry-breaking phase transition in the early universe, with formation via the Kibble mechanism \cite{Kibble:1976sj}. Under such phase transitions, vacuum configurations are uncorrelated beyond the Hubble scale and the resulting pattern of symmetry breaking results in a network of topological defects established over large scales.

In this Letter we describe a novel scenario for the formation of string networks. This scenario starts with a population 
of isolated small loops in the background of a kinating modulus field whose vev controls the tension of the strings (for example, the volume modulus in string compactifications). If the string tension decreases with time, such loops grow in physical size. If the tension decreases sufficiently rapidly, initially small loops grow faster than the scale factor and, provided the epoch lasts long enough, find each other, percolate and form a network.  

Previous papers considering strings with time-dependent tensions include 
\cite{Yamaguchi:2005gp, Ichikawa:2006rw, Cheng:2008ma, Sadeghi:2009wx, Wang:2012naa, Emond:2021vts}; see also \cite{Srivastava:2024yxf} for a recent discussion of percolating strings arising from plasma flow and primordial black holes, and \cite{Copeland:1998na} describing the evolution of a network of cosmic string loops of fixed tension.

\section{Initial Conditions and Equations of Motion}

Our scenario assumes an initial population of isolated, small loops. There are various ways these could be created -- for example, quantum nucleation from the vacuum in a time-dependent background, evaporation of ultra-small primordial black holes or as products from annihilation of stringy objects (such as brane/antibrane pairs) at the end of inflation. We leave a detailed analysis of scenarios for their origin for future work.

Our focus here is on the evolution and growth of such small loops.
Normally, they are regarded as irrelevant as they would rapidly decay from gravitational wave emission. We show that with time-dependent tensions, this no longer holds.  

The action for a Nambu-Goto string whose tension $\mu$ depends on the background space-time coordinates ($x^{\nu}$ where $\nu=0,1,2,3$) is
\begin{equation}
    S_{\rm NG} = - \int d^2 \xi \hspace{0.25 em} \mu(x^{\nu}) \sqrt{-\gamma},
\end{equation}
where $\xi^a$ parametrise worldsheet coordinates (with $a = 0,1$) and $\gamma = {\rm det } \, \gamma_{ab}$ where 
\begin{equation}
    \gamma_{ab} = g_{\alpha \beta} \frac{\partial x^{\alpha}}{\partial \xi^a} \frac{\partial x^{\beta}}{\partial \xi^b} \, .
\end{equation}
The equations of motion (also derived in \cite{Emond:2021vts}) are:
\begin{equation}\label{EOMnogf}
    {x^\nu_{,a}}^{;a} + \Gamma^\nu_{\beta \rho} (g) \hspace{0.25 em}\gamma^{ad} x^\beta_{,d} x^\rho_{,a} +\frac{\mu_{,\rho}}{\mu} \gamma^{ab} x^\rho_{,a} x^\nu_{,b} - \frac{\mu^{,\nu}}{\mu}= 0,
\end{equation}
where $x^\nu_{,b} \equiv \frac{\partial x^{\nu}}{\partial \xi^b}$,
$\Gamma^\nu_{\beta \rho} (g)$ denote the spacetime Christoffel symbols and 
\begin{equation}
    {x^\nu_{,a}}^{;a} \equiv \frac{1}{\sqrt{-\gamma}} \partial_a (\sqrt{-\gamma} \gamma^{ab} x^\nu_{,b}).
\end{equation}
For constant $\mu$, these reduce to the ordinary string equations of motion \cite{Vilenkin:2000jqa}.

We now specialise to an FLRW spacetime metric and assume that the tension only depends on time. We also impose the standard worldsheet gauge conditions, identifying worldsheet time with spacetime time $\xi^0 = x^0$  and applying the transversality condition $\dot{\vec{x}} \cdot \vec{x}' =0 $ (such that $\gamma_{01} = \gamma_{10} = 0$), where dots denote derivatives with respect to $t$ and primes denote derivatives with respect to the spatial worldsheet coordinate $\sigma$. 

The worldsheet metric then takes the form
\begin{equation}
    (\gamma_{ab}) = \begin{pmatrix}
        1- a^2 \dot{\vec{x}}^2 & 0\\
        0 & -a^2 \vec{x}'^2 
    \end{pmatrix}.
\end{equation}
It is useful to define the function $\varepsilon$, invariant under spacetime diffeomorphisms, as follows, 
\begin{equation}\label{defvarepsilon}
    \varepsilon (t,\sigma) \equiv \sqrt{\frac{-x'^2}{\dot{x}^2}} = \sqrt{\frac{a^2\vec{x}'^2}{1-a^2\dot{\vec{x}}^2}}.
\end{equation}
Eq.(\ref{EOMnogf}) then takes the form: 
\begin{eqnarray}
    \nu = 0 \hspace{0.25 em}  & : & \hspace{0.5 em} \frac{\dot{\varepsilon}}{\varepsilon} = \frac{\dot{a}}{a} (1-2 a^2 \dot{\vec{x}}^2) - \frac{\dot{\mu}}{\mu} a^2 \dot{\vec{x}}^2, \\ \label{Constraint}
   \nu = i \hspace{0.25 em}  & : &  \hspace{0.5 em} \ddot{\vec{x}} - \varepsilon^{-1} \left(\varepsilon^{-1} \vec{x}'\right)' + \left(\frac{\dot{\varepsilon}}{\varepsilon}+2\frac{\dot{a}}{a} + \frac{\dot{\mu}}{\mu}\right)\dot{\vec{x}} = 0. \label{EOM}
\end{eqnarray}

We focus on the case of circular closed strings, with periodic boundary conditions $\vec{x}(t,\sigma) = \vec{x}(t,\sigma + 2\pi)$.
With this ansatz, the string evolution can be written as 
\begin{equation}
    \vec{x} (t,\sigma) = R(t) \vec{u}(\sigma).
\end{equation}
For simplicity, we restrict to a circular loop in the $z$-plane with $\vec{u}(\sigma) = (\sin \sigma , \cos \sigma, 0)$. Substituting this ansatz into the equations of motion, we get
\begin{equation}
\frac{\dot{\varepsilon}}{\varepsilon} = H - a^2 \dot{R}^2 \left(2 H +  \frac{\dot{\mu}}{\mu} \right), \label{RConstraint}
\end{equation}
\begin{equation}
\ddot{R} + H\dot{R} + \varepsilon^{-2} R + \left(2H + \frac{\dot{\mu}}{\mu}\right)(1-a^2\dot{R}^2)\dot{R} =0, \label{REOM}
\end{equation}
where $H = \dot{a}/a$ denotes the Hubble parameter. 

The definition of the parameter $\varepsilon$ reveals its physical interpretation. An oscillating loop has zero velocity at the point of maximum amplitude and so 
\begin{align}
    \varepsilon = \sqrt{\frac{a^2R^2}{(1-a^2\dot{R}^2)}} \equiv a R_{{\rm max}}
\end{align}
represents the maximum (physical) radius of the loop at each oscillation.

Small loops (much smaller than the Hubble scale) oscillate rapidly with time-averaging giving $\langle a^2 \dot{R}^2 \rangle= 1/2$. In the standard case of  $\dot{\mu} =0$, $\frac{\dot{\varepsilon}}{\varepsilon} = 0$ and it follows from Eq.(\ref{RConstraint}) that such loops remain at constant physical size while the scale factor grows (so they shrink in comoving coordinates).

The right-hand side of Eq.(\ref{RConstraint}) contains the most important qualitative feature of the loop evolution with a time-dependent tension. If 
\begin{equation}
   2H + \frac{\dot{\mu}}{\mu} = 0, 
\end{equation}
then the physical loop radius expands at exactly the same rate as the scale factor; loops neither grow nor shrink in comoving coordinates. As $a^2 \dot{R}^2 > 0$, it follows that the condition for a loop to grow in comoving coordinates is
\begin{equation}
\label{PercCon}
2H + \frac{\dot{\mu}}{\mu} < 0.
\end{equation}  

For small, rapidly oscillating loops, we can time-average $\langle a^2 \dot{R}^2 \rangle= 1/2$
in Eq.(\ref{RConstraint}) to obtain
\begin{equation}
\frac{\dot{\varepsilon}}{\varepsilon} = - \frac{1}{2}\frac{\dot{\mu}}{\mu}.
\end{equation}
In terms of the physical length $L = 2 \pi \epsilon$, we have
\begin{equation}
    L(t) = L_i \sqrt{\frac{G\mu_i}{G\mu(t)}}. \label{lengthgrows}
\end{equation}
where $L_i$ is the initial loop length and $\mu_i$ the initial loop tension.
As we will see, in volume modulus kination $\vert \frac{\dot{\mu}}{\mu} \vert = 3H$  and thus, for loops well inside the horizon the frequency of oscillation is fast compared to the Hubble time, $\omega \sim 1/\varepsilon \gg H \sim \vert \frac{\dot{\mu}}{\mu} \vert $, justifying the time-averaging $\langle a^2 \dot{R}^2 \rangle= 1/2$. 

\section{Kination in String Theory}

What are the optimal conditions for the growth equation 
\begin{equation}
\label{PercolationEq}
  2H + \frac{\dot{\mu}}{\mu} < 0  
\end{equation}
to be satisfied? It is clear that this prefers (a) $H$ to be as small as possible and (b) the rate of change of $\mu$ to be as fast as possible. 
As to point (a), $H(t)$ is always set by the expansion rate. Compared to other fluids, the slowest expansion rate is obtained for kination where  the energy density is dominated by the kinetic energy of a rolling scalar (with $a(t) \sim t^{1/3}$ compared to e.g. $a(t) \sim t^{1/2}$ for radiation). As to point (b), in string theory, all physical scales (including string tensions) arise as expectation values of scalar fields $\Phi$ (moduli). $\vert \frac{\dot{\mu}(\Phi)}{\mu(\Phi)} \vert$ is maximised when the field $\Phi$ rolls as fast as possible; the limiting case is that of kination, when the entire energy density of the universe lies in a rolling scalar field.

This suggests that a kination environment (see \cite{Gouttenoire:2021jhk} for a recent general review of kination) gives the best opportunity to satisfy Eq.(\ref{PercolationEq}).
In the 4d Einstein frame with constant $M_P$,
the requirement that $\frac{\dot{\mu}}{\mu}$ be negative implies that the fields roll in the direction of decreasing tension. In string theory, this direction is towards the asymptotic boundaries of moduli space: in the strict asymptotic limit, all scales vanish compared to $M_P$.

As mentioned earlier, during a kination epoch, the universe is dominated by the kinetic energy of the rolling modulus,
\begin{equation} \label{eq:PhiKination}
\Phi = \Phi_i + \sqrt{\frac{2}{3}} M_P \ln \left( \frac{t}{t_i} \right),
\end{equation}
with $\Phi_i$ the initial vev at time $t_i$.
We focus on the case where $\Phi$ is the volume modulus of the compactification evolving towards large compactifiction radii. This direction is well motivated: the effective field theory becomes better controlled at large radii. Such cosmologies are also appealing from both phenomenological and formal perspectives \cite{Apers:2024ffe}.

Specialising to IIB models where moduli stabilisation is best understood, 
the canonically normalised modulus $\Phi$ relates to the compactification volume as $\Phi \sim \sqrt \frac{2}{3} \ln \mathcal{V}$ and so the volume evolves as $\mathcal{V} \propto t$ during this epoch (see \cite{Conlon:2022pnx} and \cite{Apers:2022cyl} for more detailed discussions). 

The fundamental string scale relates to the 4d Planck scale as 
\begin{equation}
    m_s \sim \frac{M_P}{\sqrt{\mathcal{V}}}.
\end{equation}
In string compactifications, all scales are tied to the fundamental scale $m_s$. As the volume increases, the string scale (in the 4d Einstein frame) decreases. In particular, the tension of string-like objects in the 4d theory decreases.

The most obvious string-like objects in string theory 
are fundamental strings, with a tension 
\begin{equation}
G \mu \sim m_s^2.    
\end{equation} 
During a kination epoch this tension behaves as
\begin{equation}
G \mu \sim t^{-1},
\end{equation}
and so 
\begin{equation}
2H + \frac{\dot{\mu}}{\mu} = -H,
\end{equation}
satisfying the growth condition Eq.(\ref{PercolationEq}). 

Note there is one further type of string present as a degree of freedom  in IIB compactifications. This is the axionic string associated to the volume modulus (for a recent review of axions and axion strings in string theory see \cite{Reece:2024wrn}). The K\"ahler potential $K = - 3 \ln (T + \bar{T})$, with $T = \tau_b + ia_b$, gives the Lagrangian
\begin{equation}
\mathcal{L} = \frac{3}{4\tau_b^2} \partial_{\mu} \tau_b \partial^{\mu} \tau_b + \frac{3}{4\tau_b^2} \partial_{\mu} a_b \partial^{\mu} a_b,
\end{equation}
from which it follows that the volume axion $a_b$ has decay constant $f_a \sim \tau_b^{-1} \sim \mathcal{V}^{-2/3}$. The presence of this axion implies the existence of associated axionic strings in the spectrum.

If the tension were given by $G\mu \sim f_a^2$, then during the kination epoch we would have $G\mu \sim M_P^2 \mathcal{V}^{-4/3} \sim M_P^2 t^{-4/3} $, falling off more rapidly than for fundamental strings.
However, the cores of such stringy axionic strings tend to involve wrapped branes (e.g. see \cite{Marchesano:2022axe}) such as a D3 brane wrapped on an internal 2-cycle to create a string in the non-compact dimensions. A string created from a D3 brane wrapped on an internal large 2-cycle has a tension $G \mu \sim R^2 m_s^2$ and so evolves as $G \mu \sim t^{-2/3}$ during a kination epoch. While the tension does decrease, such strings are not in the percolation regime as $2H + \frac{\dot{\mu}}{\mu} = 0$.

We therefore focus on fundamental strings, for which the physical length of a loop grows as
\begin{equation}
L(t) = L_i \left( \frac{t}{t_i} \right)^{1/2}.
\end{equation}
As the scale factor grows as $a(t) \sim t^{1/3}$, in comoving coordinates the radius grows as 
\begin{equation} \label{eq:RmaxComoving}
R_{{\rm max}}(t) = R_{{\rm max},i} \left( \frac{t}{t_i} \right)^{1/6}.
\end{equation}
Note that the decreasing tension makes the loops grow faster than the scale factor but not faster than the Hubble horizon. As a result, initially subhorizon loops become more and more subhorizon throughout their evolution and the fast oscillations will remain fast (indeed, becoming even quicker in comparison to the Hubble time).

\section{Gravitational Wave Emission}

Oscillating loops radiate energy and so shrink due to gravitational wave (GW) emission. It is important to check that this effect does not dominate the growth from the decreasing tension. The rate of power loss from a loop from GW emission is written
\begin{equation}
\label{powgravw}
    P_{\rm GW} = \Gamma G \mu^2,
\end{equation}
where $\Gamma$ is a numerical factor that depends on the precise loop configuration. Nambu-Goto simulations suggest $\Gamma \sim 50 - 75$~\cite{Vilenkin:2000jqa}. 

In the case of a time-dependent tension, additional terms are expected due to the change in amplitude. However, for loops well inside the horizon the oscillation rate is much faster than the rate of change in amplitude. Therefore, the main contribution to the gravitational wave emission comes from the oscillatory behaviour. Additionally, we expect a modification of the dimensionless coefficient $\Gamma$ as the loops follow a different trajectory. However, we can expect this to be a similar order of magnitude, or at least not to differ by several, and thus we will approximate the power 
emitted by Eq. \eqref{powgravw} above, where the tension is now a function of time.

It follows that the  order-of-magnitude  lifetime of a loop of length $\lambda l_s$ and mass $\lambda l_s \mu$ is (using $\mu \sim m_s^2 = l_s^{-2}$)
\begin{equation}
    \tau_{\rm GW} \sim \frac{\lambda l_s}{\Gamma G \mu} = \frac{8 \pi \lambda}{\Gamma} \frac{M_P^2}{m_s^3}.
\end{equation}
If $\tau_{\rm GW} \ll H^{-1}$, GW emission dominates the string dynamics over the effects of decreasing tension. Conversely, if $\tau_{\rm GW} \gg H^{-1}$ then the GW emission is negligible compared to the effects of the decreasing tension.

During the epoch of volume modulus kination, the background has
\begin{equation}
{\rm Energy} = \frac{\dot{\Phi}^2}{2} = \frac{M_P^2}{3} \frac{1}{t^2}.
\end{equation}
As $\mathcal{V} \propto t$ during volume modulus kination and $m_s^2 \sim \frac{M_P^2}{\mathcal{V}}$, it follows that the universe's energy density during this epoch satisfies
\begin{equation}
\rho_{\rm kin} = A \times m_s^4,
\end{equation}
with $A$ a constant, and so remains at a fixed ratio relative to the string scale energy density $m_s^4$. During kination, 
$
H^{-1} = \sqrt{\frac{3}{A}} \frac{M_P}{m_s^2}
$
and so
\begin{equation} \label{eq:tauGW}
\tau_{\rm GW} \sim \frac{8 \pi \lambda}{\Gamma} \sqrt{\frac{A}{3}}  \frac{M_P}{m_s} H^{-1} .
\end{equation}
Provided $A$ is not too small, the factor $M_P/m_s \gg 1$ implies that $\tau_{\rm GW} > H^{-1}$ and the effect of varying tension dominates over the effects of gravitational wave emission (note also that $M_P/m_s$ grows during kination). If the initial inflationary potential (which transitions into kination) arose from stringy objects such as D/$\bar{\rm D}$ brane-antibrane pairs, we would expect $A \sim 1$.

Furthermore, we can see from \eqref{eq:tauGW} that the most critical point is at the start: if gravitational wave emission does not dominate at the beginning, it will not do so later on. Due to the decreasing tension and the growth in length during kination, the ratio $\tau_{\rm GW}/(H^{-1})$ increases with time as $M_P/m_s \sim t$.

 Gravitational wave emission is a universal decay channel but there may be additional more model-dependent channels (for example, KK modes or light scalars such as axions or moduli), which must be checked in any individual model to ensure that they do not dominate over the effects of the decreasing tension. 
On dimensionful grounds, one expects a similar emitted power as Eq.(\ref{powgravw}) for other radiated modes with $M_P$-suppressed couplings. 

Moreover, as in any model of phenomenologically relevant cosmic superstrings, the underlying compactification must be such that long superstrings are stable against immediate fragmentation \cite{Witten:1985fp} (see e.g the discussion in \cite{Copeland:2003bj}).

In analogy with Abelian-Higgs cosmic string loops, which in numerical simulations efficiently decay to heavy degrees of freedom \cite{Hindmarsh:2021mnl} unless their radius is much larger than the inverse mass of the radial mode \cite{Matsunami:2019fss}, it may also be that the loops in our scenario have to be produced with initial radius somewhat larger than $m_s^{-1}$ in order that decays to string-scale degrees of freedom are negligible. 

\section{Percolation and the String Tension Today}

During kination, the string tension decreases fast enough that loops grow and, given time, will percolate. However, kination regimes are unstable: the energy density falls as $\rho_{\rm kin} \sim a(t)^{-6}$ and so sources of radiation will, over time, catch up with the kinetic energy. Indeed, one source for such  radiation is the gravitational wave emission from the oscillating loops.

As radiation catches up, the system of a scalar rolling down an exponential potential transitions into a brief period of radiation domination, before settling into a tracker solution that guides the field to the final vacuum. 

During the tracker solution on a potential $V(\Phi) = V_0 e^{-\lambda \Phi}$ (which needs $\lambda > \sqrt{6}$), the evolution of the volume modulus is  \cite{Apers:2024ffe}
\begin{equation}
\Phi = \Phi_0 +  \frac{2}{\lambda} M_P \ln \left( \frac{t}{t_0} \right),
\end{equation}
where $\Phi_0,~t_0$ are the value of the volume modulus and time when the tracker regime begins,
whereas the scale factor tracks the fluid, $a(t) \sim t^{1/2}$ for radiation. It follows that for fundamental strings during the tracker epoch $\mu \sim t^{-\sqrt{6}/\lambda}$ and so, while the tension continues to decrease, loops now grow slower than the scale factor and the percolation requirement Eq.(\ref{PercCon}) is no longer satisfied.

As the field settles down in the final vacuum, the volume is fixed and the string tension ceases to evolve. The tension then remains constant until the present day, with a value (for fundamental strings) of
\begin{equation}
    G \mu_{\rm now} \sim \langle \mathcal{V}_{\rm now} \rangle^{-1} \, \, \left( = \frac{l_s^6}{{\rm Vol}_{\rm CY}} \right).
\end{equation}
where ${\rm Vol}_{\rm CY}$ is the volume of the associated compactified Calabi-Yau space. The Large Volume Scenario (LVS) \cite{Balasubramanian:2005zx} is the most studied scenario for stabilised string vacua at exponentially large values of the volume, $\langle \mathcal{V}_{\rm now} \rangle \gg 1$. In this context, values around $\langle \mathcal{V}_{\rm  now} \rangle \sim 10^{10}$ are phenomenologically appealing (based both on stabilising the electroweak hierarchy via supersymmetry, $M_{\rm soft} \sim M_P \langle \mathcal{V}_{\rm now} \rangle^{-3/2}$ and also on ensuring that decays of the lightest volume modulus occur prior to nucleosynthesis, which requires $m_{\Phi,{\rm now}} \sim M_P \langle \mathcal{V}_{\rm now} \rangle^{-3/2} \gtrsim 30 \, {\rm TeV}$)\cite{Blumenhagen:2009gk}.

If strings succeed in percolating, they interconnect with neighbouring strings and will form a conventional string network involving long strings that is expected to enter the scaling regime. We leave to future work a detailed study of the formation of the network and the subsequent density of long strings. Such a string network, built from fundamental strings, would then survive to today with a tension $G\mu \sim \langle \mathcal{V}_{\rm now} \rangle^{-1} \sim 10^{-10}$ and a fractional energy density $\Omega_{\rm strings} \sim G\mu \sim 10^{-10}$.

Whether string loops percolate before the end of kination depends on their typical length and number density at the start of kination as well as the duration of kination. Suppose that at the start of kination  the Hubble parameter is $H_i$, the distribution of loop lengths is sharply peaked at $\lambda l_s$ and the number of loops per Hubble patch is $N_H$ (such that the typical inter-string spacing is of order $N_H^{-1/3} H_{i}^{-1}$). Using Eq.(\ref{eq:RmaxComoving}) and the definition of $A$ immediately above Eq.(\ref{eq:tauGW}), percolation occurs provided
\begin{equation} \label{eq:perc}
\frac{t_f}{t_i} \gtrsim \frac{1}{\lambda^6 N_H^2} \left(\frac{3}{A}\right)^3 \left(\frac{M_P}{m_{s,{\rm init}}}\right)^6,
\end{equation}
where $t_f$ is the time at the end of kination and $m_{s,{\rm init}}$ refers to the string scale at the start of kination.
If kination starts soon after the end of inflation and inflation comes from stringy physics at a high scale (slightly below current constraints from tensor perturbations), a reasonable expectation is that $M_P/m_{s,{\rm init}} \sim 10^2$. 
Fixing a final value of $\mathcal{V}\sim 10^{10}$ and demanding that kination ends before the volume modulus reaches its 
eventual minimum, Eq.(\ref{eq:PhiKination}) implies that $t_f/t_i\lesssim 10^{10}$ (saturated for the rather extreme case that the initial volume is close to unity in string units). Consequently, from Eq.(\ref{eq:perc}), for a long but plausible era of kination, percolation occurs provided that there are at least a few loops of length somewhat larger than $l_s$ per Hubble patch at the start of kination.

It might also be interesting to consider rare string loops nucleated during inflation \cite{Basu:1991ig}. In this case, the initial distribution of loop lengths is peaked around the Hubble length at the end of inflation, and percolation is possible even if the number of loops per Hubble patch at the end of inflation $N_H\ll 1$. In particular, again assuming kination starts soon after the end of inflation, percolation occurs provided $t_f/t_i\gtrsim N_H^{-2}$.

Gravitational radiation emitted by the loops acts as seed radiation that grows with respect to the kinating background; once this catches up, the kination period will end. Defining $a_r$ as the value of the scale factor at that time such that $\rho_{\rm GW} (a_r/a_i)^{-4} = \rho_{\rm kin} (a_r/a_i)^{-6}$, the condition of the loops to find each other and percolate before is given by 
\begin{equation}
    \frac{t_p}{t_i} \lesssim \left(\frac{24 \pi}{N_H \Gamma} \right)^{3/2}  \left(\frac{ M_P}{m_{s,{\rm init}}}\right)^{6},
\end{equation}
where $t_p$ is the time percolation happens, and we have used Eq.\eqref{eq:tauGW}  together with the definition of $A$ immediately above. We have also taken $n_{{\rm loops},i} = N_H H_i^3$. Such radiation can play a useful role in beginning the tracker era prior to overshoot.

\section{Conclusion}

This Letter describes a new mechanism for producing a cosmic (super) string network, starting from initial conditions of a population of small isolated loops. If the string tension decreases with time, the physical size of the loops grow. If the decrease is rapid enough, the loops grow faster than the scale factor; they find each other, interconnect, percolate and form a network. The key condition for this to be possible is Eq.\eqref{PercCon}, which is satisfied by loops of fundamental superstrings in the background of a kinating volume modulus evolving towards the asymptotic region of moduli space. This is an exciting new possibility, which does not rely on phase transitions, for the formation of cosmic string networks in which small loops not much larger than the string scale may grow into macroscopic long strings.

\begin{acknowledgments}
{\bf Acknowledgements}
JC acknowledges support from the STFC consolidated grants ST/T000864/1 and ST/X000761/1 and is also a member of the COST Action COSMIC WISPers CA21106, supported by COST (European Cooperation in Science and Technology). EJC acknowledges support from the STFC Consolidated Grant [ST/T000732/1] and also from a Leverhulme Research Fellowship [RF- 2021 312]. EH acknowledges the UK Science and Technology Facilities Council for support through the Quantum Sensors for the Hidden Sector collaboration under the grant ST/T006145/1 and UK Research and Innovation Future Leader Fellowship MR/V024566/1. NSG acknowledges support from the Oxford-Berman Graduate Scholarship jointly funded by the Clarendon Fund and the Rudolf Peierls Centre for Theoretical Physics Studentship. EJC would like to thank the members of the Centre for their kind hospitality during the period this work was being undertaken.

\end{acknowledgments}

\appendix

\nocite{*}

\bibliography{apssamp}

\end{document}